\documentclass{LMCS}

\usepackage{amssymb,amsmath,latexsym,hyperref}

\newtheorem{theorem}{Theorem}[section]

\newtheorem{definition}[theorem]{Definition}

\newtheorem{example}[theorem]{Example}

\newcommand{\tup}[1]{\mbox{$\overline {#1}$}}
\newcommand{\A}{\mathcal A}
\newcommand{\B}{\mathcal B}
\newcommand{\C}{\mathcal C}
\newcommand{\D}{\mathcal D}

\newcommand{\F}{\mathcal F}
\newcommand{\G}{\mathcal G}

\newcommand{\J}{\mathcal J}
\newcommand{\K}{\mathcal K}
\def\L{\mathcal L}
\newcommand{\M}{\mathcal M}

\newcommand{\R}{\mathcal R}

\newcommand{\U}{\mathcal U}

\newcommand{\con}{\otimes}

\newcommand{\N}{{\mathbb N}}

\newcommand{\Z}{{\mathbb Z}}

\newcommand{\st}{\ \vert \ }

\def\Fraisse{Fra\"{\i}ss\'{e} }
\newcommand{\1}{{\bf 1}}
\newcommand{\0}{{\bf 0}}

\def\doi{3 (2:2) 2007}
\lmcsheading%
{\doi}
{1--18}
{}
{}
{Sep.~15, 2006}
{Apr.~26, 2007}
{}   

\begin{document}

\title{Automatic Structures: Richness and Limitations }
\author[B.~Khoussainov]{Bakhadyr Khoussainov\rsuper a}
\address{{\lsuper{a,b,c}}Department of Computer Science, University of Auckland, New Zealand}
\email{\{bmk,andre,rubin\}@cs.auckland.ac.nz}
\thanks{{\lsuper{a,b}}The first and the second authors' research was
partially supported by the Marsden Fund of New Zealand.}

\author[A.~Nies]{Andr\'e Nies\rsuper b}
\address{\vskip-6 pt}

\author[S. Rubin]{Sasha Rubin\rsuper c}
\address{\vskip-6 pt}
\thanks{{\lsuper c}S.~Rubin is supported by a New Zealand Science and Technology
Post-Doctoral Fellowship.}

\author[F.~Stephan]{Frank Stephan\rsuper d}
\address{{\lsuper d}F.~Stephan,
School of Computing and Department of Mathematics, National
University of Singapore, Republic of Singapore}
\email{fstephan@comp.nus.edu.sg}
\thanks{{\lsuper d} F.~Stephan previously worked at National ICT Australia which is funded
by the Australian Government's Department of Communications, Information
Technology and the Arts and the Australian Research Council through Backing
Australia's Ability and the ICT Centre of Excellence Program.  Currently,
F.~Stephan is supported in part by NUS grant number R252-000-212-112.}


\begin{abstract}
We study the existence of automatic presentations for various
algebraic structures. An automatic presentation of a structure is a description
of the universe of the structure by a regular set of words, and the
interpretation of the relations by synchronised automata.  Our first topic
concerns characterising classes of automatic structures. 
We supply a characterisation of the automatic Boolean algebras, and it is
proven that the free Abelian group of infinite rank, as well as certain
Fra{\"\i}ss{\'e} limits, do not have automatic presentations. In particular, the
countably infinite random graph and the random partial order do not have
automatic presentations. Furthermore, no infinite integral domain is automatic.
Our second topic is the isomorphism problem. We prove that the
complexity of the isomorphism problem for the class of all automatic structures
is $\Sigma_1^1$-complete.
\end{abstract}

\maketitle

\section{Introduction}

\noindent
Classes of infinite structures with nice algorithmic properties (such as
decidable model checking) are of increasing interest in a variety of fields of
computer science.  For instance the theory of infinite state transition systems
concerns questions of symbolic representations, model checking, specification
and verification.  Also, string query languages in databases may be captured by
(decidable) infinite string structures. In these and other areas there has
been an effort to extend the framework of finite model theory to infinite
models that have finite presentations.

Automatic structures are (usually) infinite relational structures whose domain
and atomic relations can be recognised by finite automata operating
synchronously on their input. Consequently, automatic structures have finite
presentations and are closed under first order interpretability (as well as
some of extensions of first order logic).  Moreover, the model checking problem for
automatic structures is decidable.  Hence automatic structures and tools
developed for their study are well suited to these fields of computer science,
see for instance \cite{blss}.

\mbox{From} a computability and logical point of view, automatic structures are
used to provide generic examples of structures with decidable theories, to
investigate the relationship between automata and definability, and to refine
the ideas and approaches in the theory of computable structures. This paper
investigates the problem of characterising automatic structures in algebraic,
model theoretic or logical terms.

Specifically this paper addresses two foundational problems in the theory of
automatic structures. The first is that of providing {\em structure theorems} for
classes of automatic structures. Fix a class $\C$ of structures (over a given
signature), closed under isomorphism.  For instance, $\C$ may be the class of
groups $(G,\cdot)$ or the class of linear orders $(L,\leq)$.  A structure
theorem should be able to distinguish whether a given member of $\C$ has an
automatic presentation or not (a special case of this is telling whether a
given structure is automatically presentable or not). This usually concerns the
interactions between the combinatorics of the finite automata presenting the
structures and properties of the structures themselves. The second problem,
which is related to the first, is the complexity of the {\em isomorphism
problem} for classes of automatic structures. Namely, fix a class of automatic
structures $\C$. The isomorphism problem asks, given automatic presentations of
two structures from $\C$, are the structures isomorphic?

With regard to the first problem, we provide new techniques for proving that
some foundational structures in computer science and mathematics do not have
automatic presentations.  For example, we show that the \Fraisse limits of many
classes of finite structures, such as finite partial orders or finite graphs,
do not have automatic presentations. This shows that the infinite random graph
and random partial order do not have automatic presentations.
The idea is that the finite amount of memory intrinsic to the finite automata
presenting the structure can be used to extract algebraic and model theoretic
properties (invariants) of the structure, and so used to classify such
structures up to isomorphism. This line of research has indeed been successful
in investigating automatic ordinals, linear orders, trees and Boolean algebras.
For example there is a full structure theorem for the automatically presentable
ordinals; namely, they are those strictly less than $\omega^\omega$
\cite{delhomme}.  There are also partial structure theorems saying that
automatic linear orders and automatic trees have finite Cantor-Bendixson rank
\cite{krspartial}. In this paper we provide a structure theorem for the
(infinite) automatic Boolean algebras; namely, they are those isomorphic to
finite products of the Boolean algebra of finite and co-finite subsets of $\N$.

With regard to the second problem, it is not surprising that the isomorphism
problem for the class of all automatic structures in undecidable \cite{blgr}. The
reason for the undecidability is that the configuration space of a Turing
machine, considered as a graph, is an automatic structure. Moreover, the
reachability problem in the configuration space is undecidable. Thus with some
extra work, as in \cite{bl} or \cite{ikr}, one can reduce the reachability
problem to the isomorphism problem for automatic structures. In addition, the
isomorphism problem for automatic ordinals \cite{krspartial} and Boolean
algebras (Corollary~\ref{cOr:BAdec}) is decidable. For equivalence structures
it is in $\Pi^0_1$ (though not more is known), and for configuration spaces of
Turing machines is $\Pi^0_3$-complete \cite{rubinthesis}.

Hence it is somewhat unexpected that the complexity of the isomorphism
problem for the class of automatic structures is $\Sigma_1^1$-complete. The
$\Sigma_1^1$-completeness is proved by reducing the isomorphism problem for
computable trees, known to be $\Sigma^1_1$-complete \cite{gk}, to the isomorphism
problem for automatic structures.

The two problems are related in the following way. If one has a `nice' structure
theorem for a class $\C$ of automatic structures, then one expects that the
isomorphism problem for $\C$ be computationally `reasonable'. For instance, as
corollaries of the structure theorems for automatic ordinals and for automatic
Boolean algebras, one obtains that their corresponding isomorphism problems
are decidable. In contrast, the $\Sigma_1^1$-completeness of the isomorphism
problem of the class of all automatic structures tells us that the language of
first order arithmetic is not powerful enough to give a structure theorem for the
class of all automatic structures. In other words we should not expect a `nice'
structure theorem for the class of all automatic structures.

Here is an outline of the rest of the paper. The next section is a brief
introduction to the basic definitions.  Section $3$ provides some counting
techniques sufficient to prove non-automaticity of many classical structures
such as fields, integral domains and Boolean algebras. Section $4$ provides a
technique that is used to show non-automaticity of several structures, such as
the infinite random graph and the random partial order. The last section is
devoted to proving that the isomorphism problem for automatic structures is
$\Sigma_1^1$-complete.

Finally, we note that automatic structures can be generalised in several
directions: for instance, by using finite automata on infinite strings
\cite{bl,blgr}, finite ranked trees \cite{bl,beli02}, or finite unranked trees
\cite{ln}. Although this paper only deals with the finite word case, some of
the ideas presented here should give rise to a better understanding of the
generalisations. Indeed, Delhomm{\'e}, who independently proved that the random
graph has no automatic presentation (using a similar technique as the one
presented here) has extended the technique to tree-automatic structures
\cite{delhomme}. The authors would like to thank referees for comments
on improvement of this paper. 

\section{Preliminaries}

\noindent
A thorough introduction to automatic structures can be found in \cite{bl}
and \cite{kn}.  We assume familiarity with the basics of finite automata
theory though to fix notation the necessary definitions are included.
\ A {\em finite automaton} $\A$ over an alphabet $\Sigma$ is a tuple
$(S,\iota,\Delta,F)$, where $S$ is a finite set of {\em states}, $\iota \in S$
is the {\em initial state}, $\Delta \subset S \times \Sigma \times S$ is the
{\em transition table} and $F \subset S$ is the set of {\em final states}.
A {\em computation} of $\A$ on a word $\sigma_1 \sigma_2 \dots \sigma_n$
($\sigma_i \in \Sigma$) is a sequence of states, say $q_0,q_1,\dots,q_n$, such
that $q_0 = \iota$ and $(q_i,\sigma_{i+1},q_{i+1}) \in \Delta$ for all $i \in
\{0,1,\ldots,n-1\}$.  If $q_n \in F$, then the computation is {\em successful}
and we say that automaton $\A$ {\em accepts} the word. The {\em language}
accepted by the automaton $\A$ is the set of all words accepted by $\A$. In
general, $D \subset \Sigma^{\star}$ is {\em finite automaton recognisable},
or {\em regular}, if $D$ is the language accepted by a finite automaton~$\A$.

The following definitions extends recognisability to relations of arity
$n$, called {\em synchronous $n$--tape automata}.  A synchronous $n$--tape
automaton can be thought of as a one-way Turing machine with $n$ input
tapes \cite{ei69}.  Each tape is regarded as semi-infinite having written
on it a word in the alphabet $\Sigma$ followed by an infinite succession
of blanks, $\diamond$ symbols.  The automaton starts in the initial state,
reads simultaneously the first symbol of each tape, changes state, reads
simultaneously the second symbol of each tape, changes state, etc., until
it reads a blank on each tape. The automaton then stops and accepts the
$n$--tuple of words if it is in a final state. The set of all $n$--tuples
accepted by the automaton is the relation recognised by the automaton.
Here is a definition.

\begin{definition}
Write $\Sigma_{\diamond}$ for $\Sigma \cup \{\Diamond\}$ where
$\Diamond$ is a symbol not in $\Sigma$.
The {\em convolution of a tuple} $(w_1,\cdots,w_n) \in
\Sigma^{\star n}$ is the string $\con (w_1,\cdots,w_n)$ of length $\max_i|w_i|$
over alphabet $(\Sigma_{\diamond})^n$ defined as follows.
Its $k$'th symbol is $(\sigma_1,\ldots,\sigma_n)$ where $\sigma_i$ is the
$k$'th symbol of $w_i$ if $k \leq |w_i|$ and $\diamond$ otherwise.

The {\em convolution of a relation} $R \subset \Sigma^{\star n}$ is the relation
$\con R \subset (\Sigma_{\diamond})^{n\star}$ formed as the set of convolutions
of all the tuples in $R$. That is $\con R = \{\con w \st w \in R\}$.
\end{definition}

\begin{definition} \label{dfn:n-tape} {\em An $n$--tape  automaton} on $\Sigma$
  is a finite automaton over the alphabet $(\Sigma_{\diamond})^n$.  An $n$--ary
  relation $R \subset \Sigma^{{\star}n}$ is {\em finite automaton recognisable}
  (in short FA recognisable) or {\em regular} if its convolution $\con R$ is
  recognisable by an $n$--tape automaton.
\end{definition}

\noindent
We now relate $n$--tape automata to structures.  A {\em structure} $\A$ consists
of a countable set $A$ called the {\em domain} and some relations and operations on
$A$. We may assume that $\A$ only contains relational predicates as the operations
can be replaced with their graphs.  We write $\A=(A,R_1^A, \ldots, R_k^A, \ldots )$
where $R_i^A$ is an $n_i$--ary relation on $\A$. The relation $R_i$ are sometimes
called basic or atomic relations. We assume that the function $i\rightarrow n_i$
is always a computable one.

\begin{definition} \label{dfn:automatic} A structure $\A$ is {\em automatic}
over $\Sigma$ if its domain $A \subset \Sigma^{\star}$ is finite automata
recognisable, and there is an algorithm that for each $i$ produces a finite
automaton recognising the relation $R_i^A \subset (\Sigma^{\star})^{ n_i}$.
A structure is called {\em automatic} if it is automatic over some alphabet.
If $\B$ is isomorphic to an automatic structure $\A$,
then call $\A$ an {\em automatic presentation} of $\B$ and say that
$\B$ is called {\em automatically presentable} $($over $\Sigma)$.
\end{definition}

\noindent
An example of an automatic structure is the word structure $(\{0,1\}^\star,
L,R, E,\preceq)$, where for all $x,y \in \{0,1\}^\star$,
$L(x)=x0$, $R(x)=x1$, $E(x,y)$  iff $|x|=|y|$, and $\preceq$ is the
lexicographical order. The configuration graph of any Turing machine
is another example of an automatic structure. Examples of automatically
presentable structures are $(\N, +)$, $(\N, \leq)$, $(\N, S)$,  the group
$(\Z,+)$, the order on the rationals $(Q, \leq)$, and the Boolean algebra
of finite or co-finite subsets of $\N$. Note that every finite structure is
automatically presentable. We use the following important theorem without
reference.

\begin{thm} {\rm \cite{kn}}
Let $\A$ be an automatic structure. There exists an algorithm that from a
first order definition (with possible use of the additional quantifier `there
exists infinitely many') in $\A$ of a relation $R$ produces an automaton
recognising $R$.\qed
\end{thm}

\section{Proving Non-Automaticity via Counting}

\noindent
The first technique for proving non-automaticity was presented in \cite{kn}
and later generalised in \cite{bl}. The technique is based on a pumping
argument and exhibits the interplay between finitely generated (sub) algebras
and finite automata. We briefly recall the technique for completeness.

A relation $R \subset (\Sigma^\star)^n$ is called {\em locally finite} if there exists
$k,l$ with $k+l = n$ so that for every
$\bar{a}$ (of size $k$) there are at most a finite number of $\bar{b}$ (of
size $l$) such that $(\bar{a},\bar{b}) \in R$.  We write
$R \subset (\Sigma^{\star})^{k+l}$.
For $\tup{b}=(b_1,\cdots,b_m)$,
write $b \in \tup{b}$ to mean $b=b_i$ for some $i$.

We start with the following elementary but important proposition.
\begin{prop} \label{prop:locfin}
  Let $R \subset (\Sigma^{\star})^{k+l}$ be locally finite, with $k$ and $l$ as
  in the definition above.  Suppose further that $R$ is a regular 
  relation, and that the automaton for $\con(R)$ has $p$ states.
  Then
  $$\max \{|y| \st y \in \tup{y}\} - \max \{|x| \st x \in \tup{x}\} \leq p
  $$
  for every $(\tup{x},\tup{y}) \in R$ where $\tup{x}$ has $k$ elements and
  $\tup{y}$ has $l$ elements.
\end{prop}

\begin{proof} Fix $(\tup{x},\tup{y}) \in R$ and say $x' \in \tup{x}$ has
  length $\max\{|x| \st x \in \tup{x}\}$ and say $y' \in \tup{y}$ has length
  $\max\{|y| \st y \in \tup{y}\}$. So $|y'| - |x'| > p$ implies that we can
  pump the string $\con(\tup{x},\tup{y})$ between positions $|x'|$ and $|y'|$.
  Then either the automaton for $\con(R)$ accepts a string that is not in
  $\con((\Sigma^{\star})^n)$ (because one of the components contains a
  subword of the form $\diamond \Sigma$)
  or otherwise it accepts strings of the form $\con(\tup{x},\tup{z})$ for
  infinitely many $\tup{z}$, contradicting that $R$ is locally finite.
\end{proof}

\noindent
The typical application of this proposition is to prove that certain structures
do not have automatic presentations.  Assume $\A$ is an automatic structure in
which each atomic relation $R_i$ is a graph of a function $f_i$, $i=1,\ldots,
m$.  Let $a_1,a_2, \ldots$ be a sequence of some elements of $A$ such that the
relation $\{(a_i, a_j) \st j = i + 1\}$ is regular. Consider the sequence
$G_1=\{a_1\}$, $G_{n+1}=G_n\cup\{a_{n+1}\} \cup \{f_i(\bar{a}) \st \bar{a} \in
G_n, i=1,\ldots, m\}$.  By the proposition there is a constant $p$ such that
the length of all elements in $G_n$ is bounded by $p \cdot n$. Therefore the
number of elements in $G_n$ is bounded by $2^{O(n)}$. Some combinatorial
reasoning combined with this observation can now be applied to provide examples
of structures with no automatic presentations, see \cite{bl} and \cite{kn}.
For example, the following structures have no automatic presentation:
\begin{enumerate}
  \item The free group on $k>1$ generators;
  \item The structure $(\N, \mid)$;
  \item The structure $(\N, p)$, where $p:\N^2 \rightarrow \N$ is a bijection;
  \item The term algebra generated by finitely many constants with at least one
    non-unary atomic operation\footnote{Thus, elements of the term algebra are
    all the ground terms, and the operations are defined in a natural way: the
    value of a function $f$ of arity $n$ from the language on ground terms
    $g_1,\ldots, g_n$ is $f(g_1,\ldots, g_n)$.}.
\end{enumerate}
Note that each of these structures has a decidable first order theory.

In the next sections we provide other more intricate techniques for showing
that particular structures do not have automatic presentations.  We then
apply those techniques to give a characterisation
of Boolean algebras that have automatic presentations. We also prove that
$(\mathbb{Q}^+, \times)$ has no automatic presentation and show that no
infinite integral domain (in particular no infinite field) has an
automatic presentation. We also study automaticity of some \Fraisse limits.

First we introduce a very useful property true of every automatic monoid $(M, \times)$.

\begin{lem} \label{lem:product}
  For each $s_1, \ldots, s_m  \in M$,
  $|\prod_i s_i| \le \max_i |s_i| + k \lceil \log m \rceil$, where $k$ is the
  number of states in the automaton recognising the graph of $\times$.
\end{lem}

\begin{proof} Here logarithm is to base 2 and $\lceil \log n \rceil$ is the
  least $i$ such that $2^i \ge n$.  We use induction on $m$. For $m=1$, the
  inequality becomes $|s_1| \le |s_1|$.  If $m>1$ let $m=u+v$ where $u =
  \lfloor m/2 \rfloor $. Apply Proposition~\ref{prop:locfin} to the graph of
  the monoid operation $ \times$ and elements $x=\prod_{i=1}^u s_i $ and $y=
  \prod_{i=u+1}^m s_i$. Then, by induction,
    $|\prod_i s_i| \leq k+ \max(|x|, |y|)$ which is equal to
    $$
    k + \max (\max_{1 \le i \le u}|s_i| + k \lceil \log u \rceil, \max_{u+1
    \le i \le m}|s_i| + k \lceil \log v \rceil),$$ which is at most $\max_i |s_i| + k \lceil \log m \rceil,$
since $1 +\max (\lceil \log u \rceil, \lceil \log v \rceil) \leq \lceil \log m \rceil$.
\end{proof}

\subsection{Automatic Boolean algebras} \label{BA}

\noindent
All finite Boolean algebras are automatic. Thus, in this section we deal with
infinite countable Boolean algebras only.
Our goal in this section is to give a full characterisation of infinite automatic
Boolean algebras. Our characterisation can then be applied to show that
the isomorphism problem for automatic Boolean algebras is decidable.
Compare this with the result from computable algebra that the
isomorphism problem for computable Boolean algebras is
$\Sigma_1^1$-complete \cite{gk}.

Recall that a  Boolean algebra $\B=(B, \cup, \cap, \setminus, \0,\1)$
is a structure, where $\cap$ and $\cup$
and $\setminus$ operations satisfy all the basic properties of the
set-theoretic intersection, union, and complementation operations; In $\B$
the relation $a\subseteq b \iff a\cap b=a$ is a partial order in which $\0$
is the smallest element, and $\1$ is the greatest element.  The complement
of an element $b \in B$ is $\1 \setminus b$ and is denoted by $\bar{b}$.

A linearly ordered set determines a Boolean algebra in a natural way
described as follows.  Let $\L=(L,\leq)$ be a linearly ordered set
with a least element. An interval is a subset of $L$ of the form
$[a,b)=\{x \st a\leq x <b\}$, where $a, b\in L\cup\{\infty\}$. The
{\bf interval algebra} denoted by $\B_{\L}$ is the collection of all
finite unions of intervals of $\L$, with the usual set-theoretic
operations of intersection, union and complementation. Every interval
algebra is a Boolean algebra. Moreover for every countable Boolean
algebra $\A$ there exists an interval algebra $\B_{\L}$ isomorphic to
$\A$.  We write $\L_1 \times \L_2$ for the ordered sum $\sum_{l \in
\L_2} \L_1$.

\begin{lem} \label{lem:finite-cofinite}
The interval Boolean algebras $\B_{\omega \times i}$, where
$i$ is positive integer, all have automatic presentations.
\end{lem}

\begin{proof}
The Boolean algebra $\B_{\omega}$ has an automatic
presentation. Indeed, every element $X$ of $\B_{\omega}$  can be
represented by a string that codes the characteristic function of $X$. For
example, the element $[1,3)\cup [6,10)$ can be represented by the string
$0\#0110001111$ while $\N \setminus [3,4)$ can be represented by the string
$1\#0001$. The boolean operations under this representation are regular,
hence this is an automatic presentation of $\B_{\omega}$.
Now, $\B_{\omega \times i}$ is isomorphic to the Cartesian product of $i$
copies of $\B_{\omega}$.  Automatic structures are closed under
the Cartesian product, and this completes the proof.
\end{proof}

\noindent
An {\bf atom} in a Boolean algebra is a non-zero element $a$ such that for
every $b \leq a$ we have $a=b$ or $b=\0$.
Assume that $\B$ is an automatic Boolean algebra not isomorphic to any
of the algebras $\B_{\omega \times i}$. Call two elements $a,b\in B$ {\bf
$F$-equivalent} if the element $(a\cap \bar{b}) \cup (b\cap\bar{a})$ is a
union of finitely many atoms. Factorise $\B$ with respect to the equivalence
relation. Denote the factor algebra by $\B/F$. Due to the assumption on $\B$
the algebra $\B/F$ is not finite. Call $x$ in $\B$ {\bf large} if its image in
$\B/F$ is not a finite union of atoms or $\0$. For example the element $\1$
is large in $\B$ because $\B$ is not isomorphic to $\B_{\omega \times i}$.
Call an element $x$ in $\B$ {\bf infinite} if there are
infinitely many elements below it. Say that $x$ {\bf splits} $y$, for $x,y \in
B$, if $x \cap y \neq \0$ and $\bar{x} \cap y \neq \0$.  For every large element
$l \in B$ there exists an element $x \in B$ that splits $l$ such that $x \cap l$
is large and $\bar{x} \cap l$ is infinite. Also for every infinite element $i
\in B$ there exists an element $x \in B$ that splits $i$ such that either
$x \cap i$ or $\bar{x} \cap i$ is infinite.

We are now ready to prove the following theorem characterising
infinite automatic Boolean algebras.

\begin{thm} \label{thm:BA}
An infinite Boolean algebra has an automatic presentation if and
only if it is isomorphic to $\B_{\omega \times i}$ for some
positive $i \in \N$.
\end{thm}

\begin{proof}
We first construct a sequence $T_n$ of trees and elements
$a_{\sigma}\in B$ corresponding to elements $\sigma \in T_n$ as
follows. The tree $T_n$ will be a set of binary strings closed
under prefixes. Therefore it suffices to define leaves of $T_n$.
Initially, we set $T_0=\{\lambda\}$ and $a_{\lambda}=\1$. Assume
that $T_n$ has been constructed.  By induction hypothesis we may
assume that the leaves of $T_n$ satisfy the following properties:
$(1)$ There exists at least one leaf $\sigma$ such that
$a_{\sigma}$ is large in $\B$. Call the element $a_{\sigma}$
leading in $T_n$.  $(2)$ There exist $n$ leaves $\sigma_1$,
$\ldots$, $\sigma_n$ such that each $a_{\sigma_i}$ is an infinite
element in $\B$.  Call these elements sub-leading elements. $(3)$
The number of leaves in $T_n$ is greater than or equal to
$n\cdot(n+1)/2$. $(4)$ For every pair of leaves $x,y$ of $T_n$ it
holds that $x \cap y = \0$.

For each leaf $\sigma\in T_n$ proceed as follows:
\begin{enumerate}
\item  If $\sigma$ is leading then find the first length lexicographical
 $b$ that splits $a_\sigma$ such that both $a_{\sigma}\cap b$
 and $a_{\sigma}\cap \bar{b}$ are infinite and one of them is large.
 Put $\sigma 0$ and $\sigma 1$ into $T_{n+1}$.
 Set $a_{\sigma 0}= a_{\sigma}\cap b$ and $a_{\sigma 1}=a_{\sigma}\cap \bar{b}$.
\item If $\sigma$ is a sub-leading then find the first length lexicographical
 $b$ that splits $a_\sigma$ such that one of $a_{\sigma}\cap b$ or
 $a_{\sigma}\cap \bar{b}$ is infinite.
 Put $\sigma 0$ and $\sigma 1$ into $T_{n+1}$.  Set $a_{\sigma
 0}= a_{\sigma}\cap b$ and $a_{\sigma 1}=a_{\sigma}\cap \bar{b}$.
\end{enumerate}
Thus, we have constructed the tree $T_{n+1}$ and elements $a_{\sigma}$
corresponding to the leaves of the tree. Note that the inductive hypothesis
holds for $T_{n+1}$. This completes the definition of the trees.

Lemma~\ref{lem:product} is now used a number of times to justify
the following steps. There exists a constant $c_1$ such that
$|a_{\sigma \epsilon}|\leq
 |a_{\sigma}| + c_1$ for all defined elements  $a_{\sigma}$ .  Now for every
 $n$ consider the set $X_n=\{a_{\sigma} \st \sigma \mbox{ is a leaf of } T_n\}$.
 There exists a
 constant $c_2$ such that for all $x\in X_n$ we have $|x| \leq c_2\cdot n$.
 Therefore $X_n\subset \Sigma^{c_2\cdot n}$ and the number of leaves in
 $T_n$ is greater than or equal to $n \cdot (n+1)/2$.  Now for every pair
 of elements $a,b$ in $X_n$ we have $a\cap b=\0$.  Therefore the number
 of elements of the Boolean algebra generated by the elements in $X_n$
 is $2^{|X_n|}$. Now let $Y=\{b_1,\ldots, b_k\}\subset X_n$. Consider the
 element $\cup Y=b_1\cup \ldots \cup b_k$. By Lemma~\ref{lem:product} applied
 to the binary operation $\cup$ there
 exists a constant $c_3$
 such that $|\cup Y| \leq c_3\cdot n$. This gives us a contradiction because
 the number of elements generated by elements of $X_n$ clearly exceeds the
 cardinality of $\Sigma^{O(n)}$.
\end{proof}

\begin{cor} \label{cOr:BAdec}
It is decidable whether two automatic Boolean algebras are isomorphic.
\end{cor}

\begin{proof}
  Every automatic Boolean algebra is isomorphic to the Cartesian product of $i$
  copies of $B_{\omega}$, the Boolean algebra of finite and co-finite subsets
  of $\N$. This $i$ can be obtained effectively:
  Given an FA-presentation of a structure $A$ in the signature of
  Boolean algebras, one can decide if $A$ is a Boolean algebra, and if
  so compute the largest $i$ such that $A$ models \begin{quote}
    ``there are $i$ disjoint elements each with infinitely many atoms below.''
  \end{quote}
  Thus the isomorphism problem is decidable.
\end{proof}

\subsection{Commutative monoids and Abelian groups}

\noindent
Note that for groups and monoids  the term `automatic' is used in a different
way \cite{autgrps}. So to avoid confusion we say such a structure is `FA
presented' instead of saying it is `automatic', and it is `FA presentable'
instead of `automatically presentable'.

We prove that $(\mathbb{Q}^+, \times)$, or equivalently, the free Abelian
group of rank $\omega$, is not FA presentable.

\begin{thm} \label{thm:Nmonoid}
Let $(M,\times)$ be a monoid containing $(\N, \times)$ as
a submonoid. Then $(M,\times)$ is not FA presentable.
\end{thm}

\begin{proof}
Assume for a contradiction that an FA presentation of $M$
is given. Let $a_0, a_1, \ldots$ be the prime numbers, viewed as elements
of $M $, and listed in length-lexicographical order (with respect to this
presentation of $M$). Let $r_n$ be such that $a_0, \ldots, a_{r_n-1}$
are the primes of length at most $n$. Let
\begin{center}
$F_n = \{ \prod_{i\,:\,0 \le i < r_n} a_i^{\beta_i}: 0 \le \beta_i < 2^n \}$.
\end{center}
By Lemma \ref{lem:product}, each term $a_i^{\beta_i}$ has length at most $n+k
\log \beta_i \le n(1+k)$. Again by the lemma, each product has length at
most $n(1+k) + k \log r_n$. Since all the products are distinct,
\begin{center}
$2^{n r_n} \leq |F_n| \le |\Sigma|^{(1+k)n + k \log r_n}$.
\end{center}
Thus $nr_n \le \log |\Sigma|[(1+k)n + k \log r_n]$ and $r_n \in O(\log r_n)/n$,
a contradiction because $r_n $ goes to infinity. \nolinebreak
\end{proof}

\noindent
In \cite{Nies:DescribeGroups} a stronger form is proved: if  $ (\mathbb{N},
+)^r$ is a submonoid of $M$, then  $r \le \log |\Sigma| (k+1)$, where $\Sigma$
is the alphabet and $k$ the number of states needed to recognize the graph of
the  operation of $M$.

\begin{cor} \label{cOr:Qtimes}
$(\mathbb{Q}^+, \times)$ is not FA presentable.\qed
\end{cor}

\noindent
For a prime $p$, let $\mathbb{Z}[1/p]$ be the additive group of  rationals of the
form $z/p^m$, $z$ an integer, $m \in \N$, and let $\mathbb{Z}_{p^{\infty}}$ be the Pr\"ufer
group $\mathbb{Z}[1/p]/\mathbb{Z}$. Using representations to base $p$, it is easy
to give FA presentations of these groups. Hence finite direct sums of
those groups are also FA presentable. The proof of the following uses similar
methods to the ones of Theorem \ref{thm:Nmonoid}.

\begin{thm} \label{thm:Qomega}
Let $A^{(\omega)}$ denote the direct sum of infinitely many copies of the group $A$.
The infinite direct sums $\mathbb{Z}[1/p]^{(\omega)}$ and $\mathbb{Z}_{p^{\infty}}^{(\omega)}$
are not FA presentable.
\end{thm}

\begin{proof}
The proof is similar to the proof of Theorem~\ref{thm:Nmonoid}.
Suppose  there is an FA presentation, and  the $i$-th copy of the
group in question is generated by all elements of the form
$a_i/p^m$, $i,m \in \N$, where the elements $a_0, a_1, \ldots$ are
listed in length-lexicographical order.  Define $r_n$ as before,
and  consider sums  $\sum_{i< r_n} \beta_i a_i/p^n$, where $0 \le
\beta_i < p^n$. By Proposition~\ref{prop:locfin}, the definable
operation $x \mapsto x/p$ increases the length  by at most a
constant. So, using Lemma~\ref{lem:product} each term $\beta_i
a_i/p^n$ has length at most $n+ c n + k n \log p$ for appropriate
$c \in \N$.  Thus each sum has length at most $c' n + k \log r_n$.
As there are $p^{nr_n}$ distinct sums, this yields a contradiction
as before.
\end{proof}

\subsection{Integral domains}

\noindent
In our next result we prove that no infinite integral domain is FA presentable.
The following definition and lemma will be used in the next section as well.

\begin{definition} Suppose $\D$ is a structure over alphabet $\Sigma$.
Write $D^{\leq n}$ for $D \cap \Sigma^{\leq n}$; that is the elements of $D$ of
length at most $n$.  Write $P_n(D)$ for $\{x \in \Sigma^n \st \exists z \in
\Sigma^{\star} \wedge xz \in D\}$, namely all prefixes of length $n$ of all
words in the domain.
\end{definition}

\begin{lem} \label{lem:S_n}
  If $D \subset \Sigma^{\star}$ is a regular language then
  \begin{enumerate}
    \item $|P_n(D)| \in O(|D^{\leq n}|)$ and
    \item $|D^{\leq n+k}| \in \Theta(|D^{\leq n}|)$ for every constant $k \in \N$.
  \end{enumerate}
\end{lem}

\begin{proof}
Suppose the automaton recognising $D$ has $c$ states.  Then for $x \in P_n(D)$
there exists $z \in \Sigma^{\star}$ with $|z| \leq c$ such that $xz \in D$
$(\dagger)$. If $n \geq c$ then $|P_n(D)| \leq |\Sigma|^c\times |P_{n-c}(D)|$ since
the map associating $x \in P_n(D)$ with the word consisting of the first $n-c$
letters of $x$, is $|\Sigma|^c$-to-one. But by using $(\dagger)$ we see that
$|P_{n-c}(D)| \leq |D^{\leq n}|$. So
$|P_n(D)| \leq |\Sigma|^c \times |D^{\leq n}|$ as required for the first part.

Fix $k \in \N$. The mapping associating $x \in D^{\leq n+k}$ to the prefix of $x$
of length $n$ is $|\Sigma|^k$-to-one. Hence
\begin{center}
  $|D^{\leq n+k}| \leq |\Sigma|^k \times
   |P_n(D)| \leq |\Sigma|^k \times |\Sigma|^c \times |D^{\leq n}|$.
\end{center}
Since $D^{\leq n} \subset
D^{\leq n+k}$ one has that $|D^{\leq n}| \leq |D^{\leq n+k}|$
and $|D^{\leq n+k}| \in O(|D^{\leq n}|)$. This completes the proof of the lemma.
\end{proof}

\noindent
Recall that an integral domain $(D,+,\cdot,0,1)$ is a commutative ring with
identity such that $x \cdot y =  0$ only if $x =  0$ or $y = 0$.
For example every field is an integral domain.

\begin{thm} No infinite integral domain is FA presentable.
\end{thm}

\begin{proof} Suppose that $(D,+,\cdot,0,1)$ is an infinite automatic
  integral domain.  For each $n \in \N$ recall that
  $D^{\leq n}=\{u\in D \st |u| \leq n\}$.  We claim that there exists an $x$ in
  $D$ such that for all $ a,b,a',b' \in D^{\leq n}$ the condition $a\cdot x + b = a'
  \cdot x + b'$ implies that $a=a'$ and $ b=b'$. We say  such $x$ {\bf
  separates} $D^{\leq n}$.  Indeed, assume that such an $x$ does not exist.
  Then for each $x\in D$ there exist $a,b,a',b'\in D^{\leq n}$ such that
  $a\cdot x + b = a' \cdot x + b'$ but $(a,b)\neq (a',b')$. Hence, since
  $D^{\leq n}$ is finite, there exist $a,b,a',b'\in D^{\leq n}$ such that
  $a\cdot x + b = a' \cdot x + b'$ but $(a,b)\neq (a',b')$ for infinitely many
  $x$.  Thus, for infinitely many $x$ we have $(a-a')\cdot x=b'-b$. But $a \neq
  a'$ for otherwise also $b=b'$, contrary to assumption. Also there exist
  distinct $x_1$ and $x_2$ such that $(a-a')\cdot x_1=(a-a')\cdot x_2$. Since
  $D$ is an integral domain we conclude that $x_1=x_2$ which is a
  contradiction.

For each $D^{\leq n}$ we can select the length-lexicographically first $x_n$
separating $D^{\leq n}$. Now the set $E_n=\{y\st \exists a, b \in D^{\leq n} \,[y=ax_n+b]\}$
has at
least $|D^{\leq n}|^2$ many elements. However by Proposition~\ref{prop:locfin} there exists
a constant $C$ such that $E_n \subset D^{\leq n+C}$, and by Lemma~\ref{lem:S_n} the
number of elements in $D^{\leq n+C}$ is in $O(|D^{\leq n}|)$. Thus, we have a contradiction.
The theorem is proved.
\end{proof}

\begin{cor}
No infinite field is FA presentable.\qed
\end{cor}

\section{Non-automaticity of some \Fraisse Limits}

\noindent
We briefly recall the definition of \Fraisse limit; see for instance \cite{ho}.
In this section we restrict consideration to relational structures of finite
signature, although the definition can be extended to signatures with function
symbols.  Let $K$ be a class of finite structures closed under isomorphism.  We
say that $K$ has the {\bf hereditary property (HP)} if for $\A \in K$ every
substructure of $\A$ is also in $K$. We say that $K$ has the {\bf joint
embedding property (JEP)} if for all $\A,\B\in K$ there exists $\C\in K$ such
that $\A$ and $\B$ are both embeddable into $\C$.  We say that $K$ is the {\bf
age} of a structure $\D$ if $K$ coincides with the class of all finite
substructures of $\D$. It clear that the age of a structure has HP and JEP. In
fact, it is not hard to prove that a class $K$ has $HP$ and $JEP$ if and only
if $K$ coincides with the class of all finite substructure of some structure.
However, note that non-isomorphic structures may have the same age. The
following property guarantees that a class $K$ with HP and JEP defines a unique
structure up to isomorphism:

\begin{definition} A class $K$ of finite structures has {\bf amalgamation
property (AP)} if for $\A, \B, \C\in K$ with embeddings   $e:\A\rightarrow
\B$ and $f:\A\rightarrow \C$ there are $\D\in K$ and embeddings $g:\B
\rightarrow \D$ and $h:\C\rightarrow \D$ such that $ge=hf$.  \end{definition}

\noindent
Below we cite a classical result in model theory due to \Fraisse. For this
we mention that a structure $\A$ is called {\bf ultra-homogeneous} if every
isomorphism between finite substructures of $\A$ can be extended to an
automorphism of $\A$. For the proof the reader may consult \cite[Theorem $7.1.2$]{ho}.

\begin{thm} Let $K$ be a nonempty class of finite structures which
has HP, JEP, and AP. Then there exists a up to isomorphism unique countable structure $\A$, called the
{\bf \Fraisse limit of $K$}, such that
$\A$ is ultra-homogeneous and $K$ is the age of $\A$.\qed
\end{thm}

\noindent
Now we can apply the theorem to obtain several examples of structures. In
each of these examples the classes $K$ have HP, JEP, and AP.

\begin{example} Let $K$ be the class of all finite linear orders. The
\Fraisse limit of $K$
is isomorphic to the ordering of rationals.  \end{example}

\begin{example} Let $K$ be the class of all finite linear orders with
one unary predicate. The \Fraisse limit of $K$
is isomorphic $(Q; \leq, U)$, where $\leq$ is the linear order of
rationals $Q$, and $U$ is a dense and co-dense subset of $Q$.  \end{example}

\begin{example} \label{example:random-graph} Let $K$ be the class of all finite
  graphs.  The \Fraisse limit of $K$
  is known as
  the {\bf random graph}. The following is an algebraic property that gives a
  characterisation (of the isomorphism type) of the random graph $\R=(V,E)$
  $($see {\rm \cite{ho}}$)$.  For every disjoint partition $X_1,X_2$ of every
  finite set $Y$ of vertices there exists a vertex $y$ such that for all
  $x_1\in X_1$ and $x_2\in X_2$ we have $(y,x_1)\in E$  and $(y,x_2) \not \in
  E$.
An explicit description of a random graph is the following.
The set $V$ of vertices is $\N$ and $(n,m) \in E$
if in the binary representation of $n$, the term
$2^m$ has coefficient $1$.
\end{example}

\begin{example} Let $K$ be the class of all finite structures of a given
signature $L$.  The \Fraisse limit of $K$ is
known as the {\bf random $L$-structure}.  \end{example}

\begin{example} Denote by $\K_p$ the complete graph (every pair
of vertices are connected by an edge) on $p$ vertices. For $p \geq 3$, consider
the class of all finite graphs which do not contain $K_p$ as a subgraph. The
\Fraisse limit of $K$ is known as the {\bf random
$\K_p$-free graph}.  It has the following property. For every finite $K_{p-1}$-free
subgraph, say with domain $Y$, and every disjoint partition $X_1,X_2$ of $Y$,
there exists a vertex $x$ that is edge connected to every vertex in $X_1$ and no
vertex in $X_2$.  \end{example}

\noindent
Recall that an anti-chain in a partial order is a set of pairwise
incomparable elements.  A chain in a partial order is a set in which
every pair of elements are comparable.

\begin{example} \label{random-order} Let $K$ be the class
of all finite partially ordered set. The \Fraisse limit $\U$ of $K$
is known as the {\bf random partial order}. The
following is an algebraic characterisation (of the isomorphism type) of the
random partial order $\U=(U,\leq)$ $($see {\rm \cite{ho}}$)$.
 \begin{enumerate}
\item If $Z$ is a {\em finite} anti-chain of $\U$ and $X$ and $Y$ partition
$Z$ then there exists an element $z \in U$ such that for every $x \in X$,
$z > x$ and for every $y \in Y$, element $z$ is not comparable with  $ y$.
\item If $Z$ is a {\em finite} chain of $\U$ with least element $x$ and
largest element $y$ then there exists an element $z \in U$ such that $z >
x$ and $z < y$ and $z$ is not comparable with every $v \in X \setminus \{x,y\}$.
\end{enumerate} \end{example}

\begin{example} Let $K$ be the class of all finite Boolean algebras. The
\Fraisse limit of $K$
is isomorphic to the
atomless Boolean algebra. This is the Boolean algebra that satisfies the
following property. For every non-zero element $x$ there exists a nonzero $y$
strictly below $x$ $($that is $y < x)$.  By Theorem~\ref{thm:BA} this
\Fraisse limit has no automatic presentation. \end{example}

\begin{example}  Let $K$ be the class of all finite Abelian $p$-groups. The
  \Fraisse limit of $K$ is isomorphic to $G=(\mathbb{Z}_{p^{\infty}})^{\omega}$.
\end{example}

\noindent
 Note that $G$ has no
  FA-presentation by Theorem~\ref{thm:Qomega}.

 \begin{proof} The class $K$ has HP, JEP and AP, so the \Fraisse limit exists.
 To show this \Fraisse limit is isomorphic to $G$, by \cite[Lemma 7.1.4]{ho} it suffices to
 show that  the age of $G$  is $K$ (clear) and that  $G$ is weakly homogeneous.  To do so,
 suppose $A \subseteq B$ are in $K$. We have to show that each embedding of $A$ into $G$ extends to an
 embedding of $B$. We may assume that $|B:A| =p$. Since $B$ is a direct product of cyclic groups whose order
 is a power of $p$,  either $B= A \times \mathbb{Z}_p$ or there is $x \in A$ such that $x$ is
 not divisible by $p$ in $A$, and $py=x$ for some $y \in B$. In either case we can extend
 the embedding.\end{proof}

\noindent Below are examples of \Fraisse limits that have automatic presentations.

\begin{example} The linear order of rational numbers has an automatic
presentation. In fact it is straightforward to check that $(\{0,1\}^\star \cdot
1, \preceq_{lex})$ is an automatic presentation of $\mbox{$(Q,\leq)$}$.
\end{example}

\begin{example}
Let $\mbox{$U=\{u \st u \in \{0,1\}^\star \cdot 1$, $|u|$ is even$\}$}$.
The structure $(\{0,1\}^\star \cdot 1; \preceq_{lex}, U)$ is the \Fraisse
limit for the class $K$ of all finite linear orders with one unary predicate.
\end{example}

\noindent
We now provide some methods for proving non-automaticity of structures.  These
methods are then applied to prove that some \Fraisse limits do not have
automatic presentations.

Let $\A$ be an automatic structure over the alphabet
$\Sigma$. Recall $A^{\leq n}=\{v\in A \st |v| \leq n\}$.  Let
$\Phi(x,y)$ be a two variable formula in the language of this
structure. We do not exclude that $\Phi(x,y)$ has a finite number
of parameters from the domain of the structure.  Now for each
$y\in A$ and $n\in \N$ we define the following function
$c_{n,y}^{\Phi} : A^{\leq n} \rightarrow \{0,1\}$:
\[
 c_{n,y}^{\Phi}(x)=\left\{
 \begin{array}{r@{\quad \mbox{if} \quad}l} 1 & \mbox{$\A \models
   \Phi(x,y)$; }\\ 0 & \mbox{$\A \models \neg \Phi(x,y)$.}
 \end{array} \right.
\]
We may drop the superscript $\Phi$ if there is no danger
of ambiguity. In the next theorem we count the number of functions
$c_{n,y}^{\Phi}$ using the fact that $\A$ is an automatic
structure. We will use this as a criterion for proving that a
given structure is not automatically presentable.

\begin{thm} \label{bound}
Let $\A$ be an automatic structure and $\Phi(x,y)$ a first order
formula (possibly with parameters) over the language of $\A$. Then
the number of functions $c_{n,y}^{\Phi}$ is in $O(|A^{\leq n}|)$.
\end{thm}

\begin{proof} 
It is sufficient to prove that there is a constant $k$ so that the number of
functions of the form $c_{n,y}$ with $y \in A \cap \Sigma^{>n}$ is at most
$k(|A^{\leq n}|)$; this is because the $y$'s in $A^{\leq n}$ can supply at most
$|A^{\leq n}|$ many additional functions $c_{n,y}$. 

Let $(Q,\iota,\rho,F)$ be a deterministic automaton recognising
the relation $\con \Phi=\{\con(x,y) \st \linebreak[3] \A \models \Phi(x,y)\}$.  
Fix $n \in \N$. 
We will associate with each $c_{n,y}$, where $y \in A \cap \Sigma^{>n}$, two
pieces of information; namely, a function $J_y : A^{\leq n} \rightarrow Q$ and a set
$Q_y \subset Q$ as follows.  
Let $\con (x,y) = \sigma_1 \sigma_2 \cdots  \sigma_k$, where $x \in A^{\leq n}$ and 
$\sigma_i \in (\Sigma_{\diamond})^2$. 
Then define $J_y(x):=\rho(\iota, \sigma_1 \cdots \sigma_n)$. Define $Q_y
\subset Q$ as those states $s \in Q$ such that 
$\rho(s,\sigma_{n+1}\cdots \sigma_k) \in F$.  
Note that $\sigma_{n+1} \cdots \sigma_k = \con (\lambda,z)$,
for some $z \in \Sigma^*$, and is independent of $x$.

We claim that if $(J_y,Q_y) = (J_{y'},Q_{y'})$ then $c_{n,y} = c_{n,y'}$.
So suppose that $(J_y,Q_y) = (J_{y'},Q_{y'})$, and let $x \in A^{\leq n}$ be
given. Say $\con(x,y) = \sigma_1 \cdots \sigma_k$, and $\con(x,y') = \delta_1
\cdots \delta_l$. Then 
\begin{eqnarray*}
\A \models \phi(x,y) 					& \iff & 
\rho(\iota,\con(x,y)) \in F 				\\ & \iff &
\rho(J_y(x),\sigma_{n+1} \cdots \sigma_k) \in F 	\\ & \iff & 
J_y(x) \in Q_y 						\\ & \iff & 
J_{y'}(x) \in Q_{y'} 					\\ & \iff &
\rho(J_{y'}(x),\delta_{n+1} \cdots \delta_l) \in Q_{y'} \\ & \iff &
\A \models \phi(x,y'). 					
\end{eqnarray*}
Then $c_{n,y} = c_{n,y'}$ as claimed. Thus the number of functions of the form
$c_{n,y}$ ($y \in A \cap \Sigma^{> n}$) is at most the number of distinct pairs
of the type $(J_y,Q_y)$. Now the number of $Q_y$ is at most $2^{|Q|}$, so we
concentrate on bounding the number of distinct functions $J_y$ by $O(|A^{\leq n}|)$.

Note that $J_y$ depends only on the first $n$ letters of $y$ in the sense that
for $|v|\geq n$ and $w,w' \in \Sigma^{\star}$, $J_{vw} = J_{vw'}$. Now every $y
\in A \cap \Sigma^{>n}$ can be decomposed as $y = vw$ with $|v| = n$.  But the
first part of Lemma~\ref{lem:S_n} says that there are $O(|A^{\leq n}|)$ many
such $y$, and so the result follows.
\end{proof}

\noindent
We give several applications of this theorem. First we apply the theorem
to random graphs.

\begin{cor} {\rm (independently \cite{delhomme})}  The
random graph has no automatic presentation.
\end{cor}

\begin{proof}
Let $(A,E)$ be an automatic presentation of the random graph and
let $\Phi(x,y)$ be $E(x,y)$. For every partition $X_1,X_2$ of the
set $A^{\leq n}$ of vertices there exists a vertex $y$ such that
for all $x_1\in X_1$ and $x_2\in X_2$ it holds that $(x_1,y)\in E$
and $(x_2,y) \not \in E$.  Hence, for every $n$, we have
$2^{|A^{\leq n}|}$ number of functions of type $c_{n,y}$,
contradicting Theorem~\ref{bound}. Hence the random graph has no
automatic presentation.
\end{proof}

\begin{cor} Let $\A$ be a random structure of a signature $L$,
where $L$ contains at least one non-unary symbol. Then $\A$ does not have
an automatic presentation.  \end{cor}

\begin{proof}
Let $R$ be a non-unary predicate of $L$ of arity $k$. Consider
the following formula $E(x,y)=R(x,y,a_1,\ldots, a_{k-2}),$ where $a_1,\ldots,
a_{k-2}$ are fixed constants from the domain of $\A$. It is not hard to prove that
$(A,E)$ is isomorphic to the random graph. But if $\A$ is automatically
presentable then so is the random graph $(A,E)$, hence contradicting the
previous corollary.
\end{proof}

\noindent
The next goal is to show that the random $K_p$-free graph has no automatic
presentation.  For this one needs to have a finer analysis than the one for
the random graph.  Let $\F=(V,E)$ be a finite graph.
For a vertex $v$, write $E(v)$
for the set of vertices adjacent to $v$. The {\em degree} of a vertex is the
cardinality of $E(v)$.  Write $\Delta(\F)$ for the maximum degree over all the
vertices of $F$. Call a subgraph $\G$ with no edges an {\em independent}
graph. Let
$\alpha(\F)$ be the number of vertices of a largest independent subgraph of $\F$.

$K_p$ denotes the complete graph on $p$ vertices; that is, there
is an edge between every pair of vertices. A graph is called {\em $K_p$-free}
if it has no subgraph isomorphic to $K_p$.

\begin{lem} For every $p \geq 3$, there is a polynomial $Q_p(x)$ of degree
    $p-1$ so that if $\F$ is a finite $K_p$-free graph then
    $Q_p(\alpha(\F)) \geq |F|$.
\end{lem}

\begin{proof}
  We first prove that for every finite graph $\F$,
  \begin{equation} \label{eqn:Cp}
      \alpha(\F) \geq |F|/(\Delta(\F)+1).
    \end{equation}
    Let $\G$ be an independent subgraph of $\F$ with a maximal number of
    vertices. That is, $\alpha(\F)=|\G|$.  For every $d \in G$ let $N(d)=E(d)
    \cup \{d\}$, where $E(d)$ is the set of vertices in $\F$ adjacent to $d$.
    Then since $\G$ is maximal, for every $x \in F$ there is some (not
    necessarily unique) $d \in G$ such that $x \in N(d)$. Hence $\F = \cup_{d
    \in G} N(d)$. But $|N(d)|$ equals the degree (in $\F$) of $d$ plus one, and
    so the largest cardinality amongst the $N(d)$'s is at most $\Delta(\F)+1$.
    Hence $|F| \leq |G| \times (\Delta(\F)+1)$ as required.

    The lemma is proved by induction on $p$. We will show that
    $Q_p(x) = \Sigma_{i=1}^{p-1} x^{i}$.
    For the case $p=3$ note that for every vertex $v$, the subgraph on domain
    $E(v)$ is independent.
    For otherwise if $x,y \in E(v)$ were joined by an edge then the
    subgraph of $\F$ on $\{x,y,v\}$ is $K_3$. In particular, $\alpha(\F) \geq
    \Delta(\F)$.  Combining this with Inequality (\ref{eqn:Cp}), we get
    $\alpha(\F)[\alpha(\F)+1] \geq |F|$ as required.

    For the inductive step, let $\F$ be a $K_p$-free graph with $p > 3$. For every
    vertex $v$, the set $E(v)$ is $K_{p-1}$-free for otherwise the subgraph
    of $\F$ on $E(v)
    \cup \{v\}$ has a copy of $K_p$.  Applying the induction hypothesis
    to $E(v)$ such that $|E(v)| = \Delta(\F)$, we get that $E(v)$ must have an independent set $X$ so that
    $Q_{p-1}(|X|) \geq |E(v)|$. But $X$ is also independent in $\F$ so
    $Q_{p-1}(\alpha(\F)) \geq \Delta(\F)$. Combining this with Inequality
    \ref{eqn:Cp}, we get that
    $\alpha(\F) [Q_{p-1}(\alpha(\F))+1] \geq |F|$. Hence $Q_p(\alpha(\F)) \geq
    |F|$ as required.
  \end{proof}

  \begin{cor} \label{cOr:kpfree} \index{non automaticity!random $K_p$-free
    graph}
    For $p \geq 3$, the random $K_p$-free graph is not automatically presentable.
  \end{cor}

  \begin{proof}
    Fix $p \geq 3$ and let $(D,E)$ be a copy of the random $K_p$-free graph.
    Then for every $K_{p-1}$-free subset $K \subset
    D^{\leq n}$ there exists an $x \in D$ that is connected to every vertex in $K$ and none
    in $D^{\leq n} \setminus K$. So let $\G_n$ be an independent subgraph of
    $\D^{\leq n}$ so that
    $Q_p(|G_n|) \geq |D^{\leq n}|$ as in the lemma.  So letting $\Phi(x,y)$ be $E(x,y)$,
    for a fixed $n$ the number of functions $c_{n,y}$ as $y$ varies over $D$
    is at least $2^{|G_n|}$ which is not linear in $|D^{\leq n}|$. Hence by
    Theorem~\ref{bound} the random $K_p$ graph $\D$ is not automatically presentable.
  \end{proof}

\noindent
As the fourth application we prove that the random partial
order $\U$ does not have an automatic presentation. For the proof we need the
following combinatorial result that connects the size of a finite partial order
$(B,\leq)$ with the cardinalities of its chains and anti-chains.

\begin{lem} [Dilworth] \label{lem:dilworth} Let $(B,\leq)$ be a finite
  partial order of cardinality $n$. Let $a$ be the size of largest anti-chain
  in $(B,\leq)$ and let $c$ be the size of the largest chain in $(B,\leq)$.
  Then $n \leq ac$.  \end{lem}

\begin{proof}
  For $1 \leq i \leq c$ define $X_i$ as the set of all elements $x$ such that
  the size of the largest chain in the subpartial order $(\uparrow x)=\{y \in B
  \st x \preceq y\}$ is $i$.  Then the $X_i$'s partition $B$. Moreover if $a
  \prec b$ and $b \in X_i$ then the size of the largest chain in $(\uparrow a)$
  is $>i$. Hence each $X_i$ is an anti-chain. Thus $\B$ can be partitioned into
  exactly $c$ many anti-chains. If $a$ is the size of the largest anti-chain in
  $\B$ then $n \leq ac$ as required.
\end{proof}

\begin{cor} \label{cOr:p} The random partial order
$\U=(U,\leq)$ has no automatic presentation.  \end{cor}

\begin{proof}
Recall that the random partial order has the following property
\begin{enumerate}
\item If $Z$ is a {\em finite} anti-chain of $\U$, and $X$ and $Y$ partition
$Z$, then there exists an element $z \in U$ such that $z > x$ for every $x \in X$, 
and $z$ is not comparable with $y$ for every $y \in Y$.
\item If $Z$ is a {\em finite} chain of $\U$ with least element $x$ and
largest element $y$, then there exists an element $z \in U$ such that $z >
x$ and $z < y$ and $z$ is not comparable with every $v \in X \setminus \{x,y\}$.
\end{enumerate}
Assume that $\U$ has an automatic presentation $(A,\leq)$.
The formula $\Phi(x,y)$
is $x \leq y \vee y \leq x$. Now let us take  $A^{\leq n}$.

Let $Z$ be an anti-chain of $A^{\leq n}$.  Consider a subset $X$ of $Z$. There
exists an element $y \in U$ such that for every $x \in X$, $y > x$ and for
every $x^{\prime}\not  \in Z \setminus X$, element $y$ is not comparable
with  $ x^{\prime}$. From this we conclude that 
$$ (\star) \hspace{20mm} \# \mbox{(functions of type $c_{n,y}$)} \geq 2^{|Z|}.  $$
Let $Z$ be a chain of $A^{\leq n}$ with least element $v$ and largest
element $w$. Then there exists an element $y \in U$ such that $y > v$ and $y
< w$ and $y$ is not comparable with  $x$ for every $x \in Z$. From this we
conclude that $$ (\star \star) \hspace{20mm}  
\# \mbox{(functions of type $c_{n,y}$)} \geq \binom{|Z|}{2}.  $$ 
Let $X$ be the largest anti-chain and $Y$ be the largest chain of $A^{\leq n}$
with cardinalities $a$ and $c$, respectively. Using Dilworth Lemma,
substituting $a$ for $|Z|$ in $(\star)$, and $c$ for $|Z|$ in $(\star\star)$,
one can, with a little algebra, derive a contradiction to the bound in the
statement of Theorem \ref{bound}.  
\end{proof}

\section{The Isomorphism Problem}

\noindent
The results in the previous sections give us hope that one can characterise
automatic structures for certain classes of structures, e.g. Boolean
algebras. However, in this section we prove that the isomorphism problem
is $\Sigma_1^1$-complete, thus showing that the problem is  as hard as
possible when  considering  the class of {\em all} automatic structures.
Then {\bf complexity of the isomorphism problem} for automatic structures
consists in establishing the complexity of the set $\{(\A, \B) \st \A$ and $\B$
are automatic structures and $\A$ is isomorphic to $\B\}$.

Let $\M$ be a Turing machine over input alphabet $\Sigma$. Its configuration
graph $C(\M)$ is the set of all configurations of $\M$, with
an edge from  $c$ to $d$ if $T$ can move from $c$ to $d$  in a single
transition. The following is an easy lemma:

\begin{lem} \label{lem:conf} For every Turing machine $T$ the configuration
graph $\C(T)$ is automatic.  Further, the set of all vertices with
outdegree $($indegree$)$ $0$ is FA-recognisable.\qed  \end{lem}

\begin{definition} \index{reversible|dd} A Turing machine $\R$ is {\bf
reversible} if every vertex in $C(\R)$ has both indegree and outdegree at
most one.  \end{definition}

\noindent
Now let $\R$ be a reversible Turing machine. Consider its configuration
space $C(\R)$.  The machine $\R$ can be  modified so that it only halts
in an accepting state; so, instead of halting in a rejecting state, it
loops forever.  Let $x$ be a configuration of $\R$.  Consider the sequence:
$x=x_0$, $x_1$, $x_2$, $\ldots$ such that $(x_i, x_{i+1})\in E$, where $E$
is the edge relation of the configuration  space. Call this sequence the {\bf
chain defined by $x$}.  We say that $x$ is the {\bf base} of chain $X$.  If $x$
does not have a predecessor then the chain defined by $x$ is maximal.  Since $\R$
is reversible, the configuration space $C(\R)$ is a disjoint union of maximal chains
such that each chain is either finite, or  isomorphic to $(\N,S)$, or  isomorphic
to $(\Z,S)$, where $(x,y)\in S$ iff $y=x+1$. It is known that every Turing machine
can be converted into an equivalent reversible Turing machine (see for example
\cite{be}).  Our next lemma states this fact and provides some additional
structural information about the configuration space of reversible Turing machines:

\begin{lem} A deterministic Turing machine can be converted into an
equivalent reversible Turing machine $\R$ such that every maximal chain
in $C(\R)$ is either finite or isomorphic to $(\N,S)$.  \end{lem}

\noindent
Denote by $\N^{\star}$ the set of all finite strings from $\N$. A set $T
\subset \N^{\star}$ is a {\em special tree} if $T$ is closed downward, namely $xy  \in T$ implies $x \in T$, for $x,y \in \N^{\star}$. We view these special trees  as structures of the
signature $E$, where $E(x,y)$ if and only if $y=xz$ for some $z\in \N$. Thus,
for every $x\in \N^\star$ the set $\{y \st E(x,y)\}$ can be thought as the
set of all {\em immediate successors of $x$}.

A special tree $T$  is {\em recursively enumerable}  if the set $T$ is
the domain of the function computed by a Turing machine. We will  use the following fact from
computable model theory \cite[Thm 3.2]{gk}.

\begin{lem} \label{lemma:computable-trees} The
isomorphism problem for recursively enumerable special trees   is $\Sigma_1^1$-complete. In fact, the trees can be chosen to be subtrees of $\{2n: n \in \N \}^*$.
\end{lem}

\noindent
It is clear from the proof  in \cite{gk} that the trees obtained are special (namely, subtrees of $\N^{\star}$).
By a mere change of notation one obtains subtrees of $\{2n: n \in \N \}^*$.

\begin{lem} The special tree $\N^{\star}$ has an automatic presentation $\A_1$.
\end{lem}

\begin{proof}
Consider the prefix relation $\leq_p$ on the set of all
binary strings.  The tree $\N^{\star}$ is isomorphic to the automatic
successor tree
$\A_1=(\{0,1\}^{\star}1 \cup \{\lambda\};  E_1)$, where $E_1$ is the set of all
pairs $(x,y)$ such that $y$ is the immediate $<_p$-successor of $x$.
The isomorphism is established via the  computable mapping
sending
$n_1 \ldots n_k$ to $0^{n_1}1 \ldots 0^{n_k}1$ and the root to $\lambda$.
\end{proof}

\noindent
Define $\mathcal{S} = \{0,1\}^{\star}1 \cup \{\lambda\}$.  From now
on we make the following conventions: 

\begin{enumerate} 
\item All special trees we consider will be viewed as  subsets of $(\mathcal{S}, \leq_p)$.  
\item The set of inputs for all Turing machines considered are strings from
$\mathcal{S}$.  Each $w\in \mathcal{S}$ is identified with  the initial
configuration starting from $w$.  
\item All Turing machines considered are reversible.  
\item The domains of the functions computed by the Turing machines are assumed to
be downward closed.  Thus, these domains form recursively enumerable special
trees.  
\end{enumerate}
Our goal is to transform every recursively enumerable tree $T\subset \mathcal{S}$
into an automatic structure $\A_{\R}$, where $\R$ is the reversible machine
with domain $T$. The idea is to attach computations of $\R$
to the initial configurations in $\mathcal{S}$. For all recursively enumerable trees $T_1$ and
$T_2$,  $T_1$ and $T_2$ are isomorphic if and only if the automatic
structures $\A_{{\R}_1}$ and $\A_{{\R}_2}$ are isomorphic.  To ensure this we
also have to attach infinitely many chains of each finite length to the initial
configurations, for in that case the length of a halting computation does not
make a difference.

Let $\R$ be a reversible Turing machine.  A {\bf chain} is an initial segment
of $(\N,S)$. Let $\J$ be the graph consisting of infinitely many finite chains
of every finite length. We denote by $J$ the set of vertices of $\J$, and the
bases of the chains in $\J$ by $b_1, b_2,\ldots$. The following is not hard to
prove:

\begin{lem} The graph $\J$ has an automatic
presentation.
\end{lem}

\noindent
To construct $\A_{\R}$, with every $w \in A_1$ associate a graph
$\J_w$ defined as follows. The vertex set of $\J_w$ consists of all
$\{(w,j) \st j \in J\}$ and edges between $(w,j)$ and $(w,j')$ if and only
if $(j,j')$ is an edge in $\J$. Put connecting edges from $w$ into each
$(w,b_i)$.  Let $\A_2$ be the graph consisting of $\A_1$ and
every $\J_w$ and the connecting edges.  Clearly, the graph $\A_2$ is
automatic.

To the set $A_2$ add all the configurations of the Turing machine $\R$ and all
the edges of the configuration space of $\R$. Note that each $w \in
\mathcal{S}$, which is an initial configuration of $\R$ is by Convention $2$
already in $A_2$. In addition, add a disjoint automatic copy of the graph $\J$,
and an automatic copy of infinitely many infinite chains. Call all chains added
at this stage {\bf junk chains}.  The resulting graph is denoted by $\A_{\R}$.

The proof of the following is straightforward and is omitted.

\begin{lem}
The graph $\A_{\R}$ is automatic.\qed
\end{lem}

\noindent
We summarise the structure of the graph $\A_{\R}$.  Firstly, in
$\A_{\R}$ there are infinitely many junk chains of every (finite and
infinite) length. An {\bf isolated chain} of $\A_{\R}$ is one in which
every vertex, except the base, has exactly one successor in $\A_{\R}$.  

With each element $w$ in $\A_1$ there is an associated structure $\J_w$, which
consists of infinitely many isolated chains of every finite length, and no
infinite isolated chain. Finally with every initial configuration $w$ (which is
an element of $\A_{\R}$ and belongs to the infinitely branching tree
$\A_1$) there is an isolated chain with base $w$ that is the computation path
from the configuration space of Turing machine $T$. These observations prove
the following.

\begin{lem} 
Let $w$ be an initial configuration of $\A_{\R}$.  Then $w$
is the base of infinitely many isolated chains of every finite length.  Also
the Turing machine $T$ halts on $w$ if and only if there is no infinite
isolated chain with base $w$.  In case $T$ does not halt on $w$ there is
exactly one infinite isolated chain with base $w$.\qed
\end{lem}

\begin{lem} Let $T_1$ and $T_2$ be computable trees which are domains of
reversible Turing machines $\R_1$ and $\R_2$ respectively. Then the trees $T_1$
and $T_2$ are isomorphic if and only if the automatic graphs $\A_{{\R}_1}$ and
$\A_{{\R}_2}$ are isomorphic.
\end{lem}

\begin{proof}
First note that the domain $T$ of $\R$ consists of all $w$ in $\A_{{\R}}$ that are
initial configurations and are not the base of an infinite isolated chain. Hence $T$ may
be thought of as a subgraph of $\A_{{\R}}$. Now suppose $\A_{{\R}_1}$ is isomorphic
to $\A_{{\R}_2}$ via the map $\phi$. Note that $w$ is the base of an infinite isolated chain
if and only if $\phi(w)$ is the base of an infinite isolated chain.  Hence $\phi$ is an
isomorphism between the trees $T_1$ and $T_2$ viewed as subgraphs of $\A_{{\R}_1}$
and $\A_{{\R}_2}$ respectively.

Conversely assume $T_1$ and $T_2$ are isomorphic via $\psi$ (as before we may
assume that $T_i$ is a subgraph of $\A_{{\R}_i}$).  Consider the sets $I_1$ and $I_2$
of all junk chains in $\A_{{\R}_1}$ and $\A_{{\R}_2}$. There is an isomorphism
between $I_1$ and $I_2$, since both consist of infinitely many chains of
every length (finite and infinite), independently of the chains of the configuration
graphs that are {\em not} defined by initial configurations. Therefore it suffices
to establish an isomorphism extending $\psi$ on the rest of the structure.

If $w \in T_1$ then there is an isomorphism between all the isolated chains with base $w$
and all the isolated chains with base $\psi(w)$.  Indeed each is the base of
infinitely many isolated chains of every finite length and no isolated chain of
infinite length, independently of the (possibly different) lengths of the
finite computations of $R_1$ on $w$ and $R_2$ on $\psi(w)$. Hence extend $\psi$
from the chains defined by elements of $T_1$ to the chains defined by elements
of $T_2$.

Suppose $w \in T_1$ and consider the set $S_1$ of immediate successors of $w$  that
are initial configurations but not in $T_1$. Similarly write $S_2$ for the set of
immediate successors of $\psi(w)$ that are initial configurations but not in $T_2$.
By the extra condition in Lemma  \ref{lemma:computable-trees} that the trees be subtrees of $\{2n: n \in \N \}^*$, both $S_1$ and
$S_2$ are infinite. So we may extend $\psi$ to those immediate successors by adding a bijection between $S_1$ and $S_2$.

Finally, for any initial configurations $v_1  \in \mathcal{S}-T_1$ and $v_2\in
\mathcal{S}- T_2$, there is an isomorphism between the nodes in $\mathcal{S}$
extending $v_1$ and the ones extending $v_2$. This isomorphism extends to the
attached chains (as there is one infinite isolated chain and infinitely many
isolated chains of each finite length). So we may extend $\psi$ to an
isomorphism of $\A_{{\R}_1}$ and $\A_{{\R}_2}$.  
\end{proof}

\noindent
Hence we have reduced the isomorphism problem for recursively enumerable trees
to the isomorphism problem for automatic graphs. The main result now follows.

\begin{thm}
The isomorphism problem for automatic structures is $\Sigma_1^1$-complete.\qed
\end{thm}

\bibliographystyle{plain}

\end{document}